\documentclass[
    ,final            
  ]
  {aipproc}

\layoutstyle{6x9}

\newcommand\apj{{ApJ}}%
\newcommand\aap{{A\&A}}%
\newcommand\mnras{{MNRAS}}%
\def\la{\mathrel{\hbox{\rlap{\hbox{\lower4pt\hbox{$\sim$}}}\hbox{$<$}}}}
\def\ga{\mathrel{\hbox{\rlap{\hbox{\lower4pt\hbox{$\sim$}}}\hbox{$>$}}}}

\begin{document}

\title{\emph{Swift} observations of the SFXT SAX~J1818.6-1703 in outburst}

\classification{97.80.Jp, 98.70.Qy, 98.70.Rz}
\keywords{X-rays: binaries -- X-rays: individual: SAX J1818.6--1703}

\author{P.\ Romano}{
  address={INAF-IASF Palermo, 
 Via U.\ La Malfa 153, I-90146 Palermo, Italy}
}
\author{L.\ Sidoli}{
  address={INAF-IASF Milano, 
 Via E.\ Bassini 15, I-20133 Milano, Italy}
}
\author{P.\ Esposito}{
  address={INAF-IASF Milano, 
 Via E.\ Bassini 15, I-20133 Milano, Italy}
}
\author{V.\ La Parola}{
  address={INAF-IASF Palermo, 
 Via U.\ La Malfa 153, I-90146 Palermo, Italy}
}
\author{J.A.\ Kennea}{
  address={Department of Astronomy and Astrophysics, PSU, University Park, PA 16802, USA}
}
\author{H.A.~Krimm}{
  address={NASA/Goddard Space Flight Center, Greenbelt, MD 20771, USA}
,altaddress={Universities Space Research Association, Columbia, MD, USA} 
}
\author{M.M.\ Chester}{
  address={Department of Astronomy and Astrophysics, PSU, University Park, PA 16802, USA}
}
\author{A.\ Bazzano}{
  address={INAF-IASF Roma, 
 Via Fosso del Cavaliere 100, I-00133, Roma, Italy}
}
\author{D.N.\ Burrows}{
  address={Department of Astronomy and Astrophysics, PSU, University Park, PA 16802, USA}
}
\author{N.\ Gehrels}{
  address={NASA/Goddard Space Flight Center, Greenbelt, MD 20771, USA}
}

\begin{abstract}
We present the {\it Swift} observations of the supergiant fast X-ray transient 
(SFXT) SAX~J1818.6--1703 collected during the most recent outburst, 
which occurred on May 6 2009. In particular, we present broad-band spectroscopic 
and timing analysis as well as a {\it Swift}/XRT light curve that spans more than 
two weeks of observations.  The broad-band spectral models and length of the outburst 
resemble those of the prototype of the SFXT class, XTE J1739--302, further confirming 
SAX J1818.6--1703 as a member of this class.
\end{abstract}

\maketitle


\section{Results}  

Here we summarize our main results on our campaign on SAX~J1818.6--1703,
obtained as a response of a {\it Swift} Cycle 5 GI trigger; 
further details are in \citet{Sidoli2009:sfxts_sax1818}. 

\begin{figure}
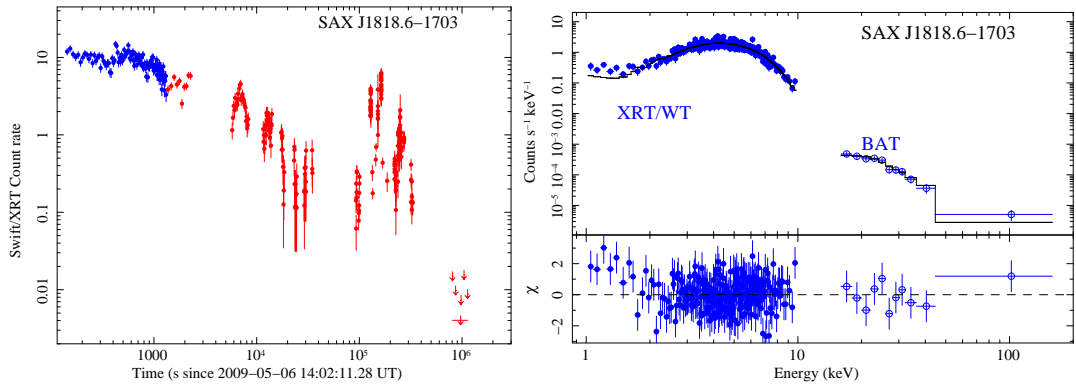

\includegraphics[width=5cm,angle=270]{romanop_box09_f1a.ps} 
\includegraphics[width=5cm,angle=270]{romanop_box09_f1b.ps} 
\vspace{1.0truecm}
\caption{{\bf Left:} {\it Swift}/XRT light curve, during the whole
campaign (2009-05-06 to 2009-05-19). The four day gap before the 3$\sigma$ upper limits
is due to the source being Moon constrained. The data
shown were collected through both AT and GI ToO
observations as well as regular ToO observations. 
The blue points were fitted simultaneously with BAT data. 
{\bf Right:} Spectroscopy of the 2009 May 6 outburst. 
Top: simultaneous XRT/WT (filled circles) and BAT
(empty circles) data fit with an absorbed Comptonization
emission (bmc model in XSPEC).
Bottom: the residuals of the fit (in units of standard deviations).
} \label{fig1}
\end{figure}

We followed the evolution of the whole May 6 outburst of SAX~J1818.6--1703 
and its declining phase with {\it Swift}/XRT. 
The source displays a dynamical range of more than 3000 (see Fig.~\ref{fig1}, left) and a
multiple-flaring behaviour as also observed in other SFXTs 
\citep{Romano2009:sfxts_paper08408}.  
Temporally resolved spectroscopy of the 
XRT data did not reveal variability in the spectral parameters within the uncertainties, 
except for the change in the blackbody radius, 
when fitting the soft X--rays with a single absorbed blackbody. We also did not find 
variability in the absorbing column density during the outburst.

We obtained, for the first time for this source, broad-band spectroscopy from soft 
to hard X--rays. 
Figure~\ref{fig1} (right) shows the {\it Swift}/XRT$+$BAT joint spectrum during the outburst 
obtained from strictly simultaneous data (only the blue XRT points in Fig.~\ref{fig1} left
were considered). 
The broad-band spectrum can be characterized by either a very high absorption, 
a flat power law ($\Gamma\sim$0.1--0.5) with a cut-off at about 7--12 keV,
or by Comptonized emission from a cold and optically thick
corona, with an electron temperature $kT_e = 5$--7\,keV, 
a hot seed photon temperature, $kT_0=1.3$--1.4\,keV, 
and an optical depth for the Comptonizing plasma $\tau=10$. 
The inferred 1--100\,keV luminosity is  $3\times 10^{36}$ erg s$^{-1}$ 
(at 2.5 kpc;  \citep[][]{Negueruela2007sax1818}).

These broad-band properties are reminiscent of the X--ray spectral shape of the 
prototype of the SFXT class, 
XTE J1739--302 \citep{Sidoli2009:sfxts_paperIII,Sidoli2009:sfxts_paperIV}: 
similarly high absorption, X--ray luminosities, energy cutoff values, and hot seed photon 
temperatures. These spectral parameters indicate the presence of a cold 
and optically thick corona. 

The observed properties of the broad band spectrum as well as the outburst 
duration of about 5 days, are quite similar to the ones observed in other SFXTs 
we have been monitoring in the last year and a half with {\it Swift} 
\citep[][ and references therein]{Romano2009:sfxts_paperV}  
and 
confirm the fact that SAX~J1818.6--1703 belongs to the class of SFXTs.

\begin{theacknowledgments}
This work was supported by contract ASI/INAF I/088/06/0 and
I/023/05/0 in Italy, by NASA contract NAS5-00136 at PSU.
\end{theacknowledgments}

\bibliographystyle{aipproc}   


\end{document}